\documentclass[nonacm,sigplan]{acmart}

\usepackage{diagbox}
\usepackage{epsfig}

\usepackage[]{hyperref}

\makeatletter
\renewcommand\subsubsection{\@startsection{subsubsection}{3}{\z@}%
  {-3.25ex\@plus -1ex \@minus -.2ex}%
  {1ex \@plus .2ex}%
  {\normalfont\normalsize\bfseries\boldmath}}
\makeatother

\begin{document}

\begin{titlepage} 

	\centering 
	
	\scshape 
	
	\vspace*{\baselineskip} 
	
	
	\rule{\textwidth}{1.6pt}\vspace*{-\baselineskip}\vspace*{2pt} 
	\rule{\textwidth}{0.4pt} 
	
	\vspace{0.75\baselineskip} 
	
	{\LARGE Achieving Consistent and Comparable CPU Evaluation} 
	
	\vspace{0.75\baselineskip} 
	
	\rule{\textwidth}{0.4pt}\vspace*{-\baselineskip}\vspace{3.2pt} 
	\rule{\textwidth}{1.6pt} 
	
	\vspace{2\baselineskip} 
	
	
	
	\vspace*{3\baselineskip} 
	
	
	Edited By
	
	\vspace{0.5\baselineskip} 
	
	{\scshape\Large Chenxi Wang\\ Lei Wang\\ Wanling Gao \\ Fanda Fan \\ Yuchen Su \\ Yutong Zhou \\ Yikang Yang \\Jianfeng Zhan\\}

	\vspace{0.5\baselineskip} 

	\vfill 
	
	
	\epsfig{file=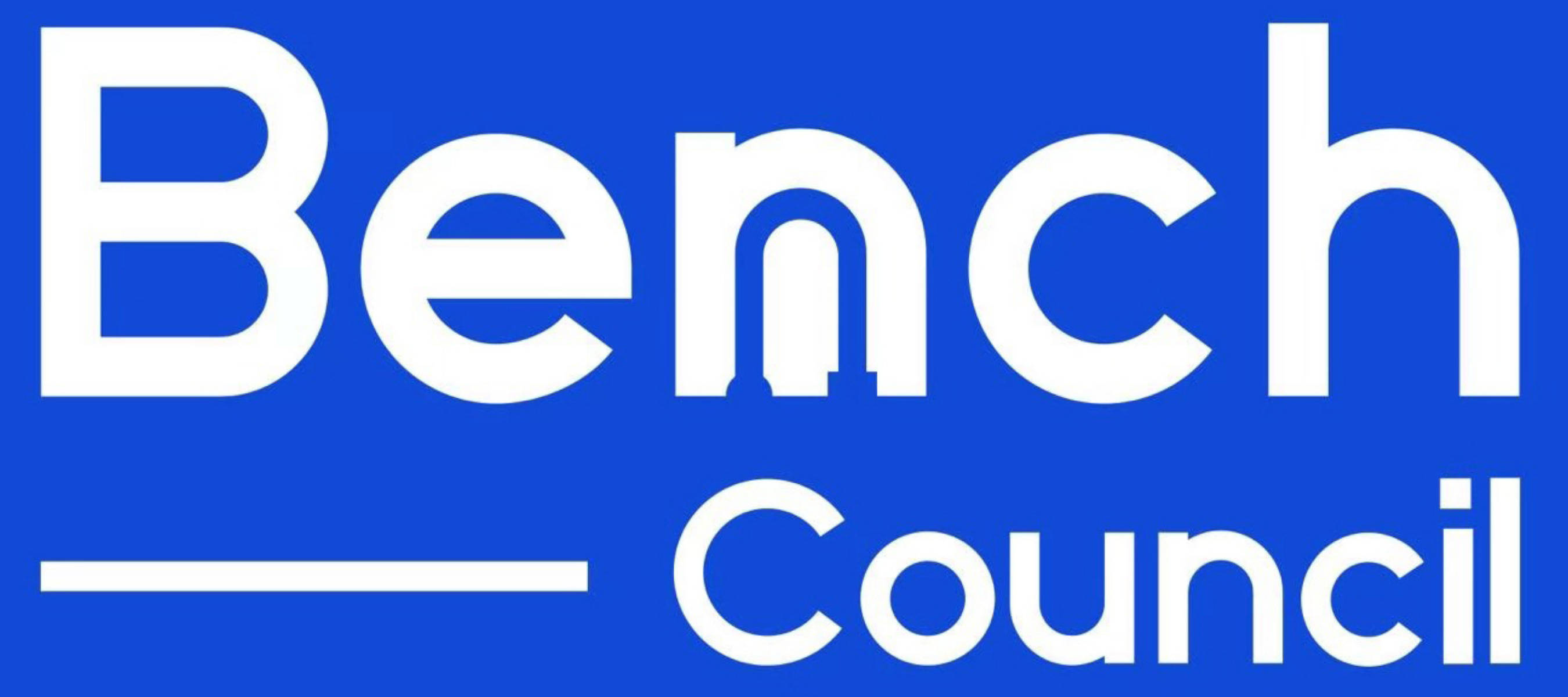,height=2cm}
	\textit{\\BenchCouncil: International Open Benchmark Council\\http://www.benchcouncil.org} 
	\vspace{5\baselineskip} 

	Technical Report No. BenchCouncil-CPU Evaluatology-2025 
	
	{\large Jul 15, 2025} 

\end{titlepage}

\title{Achieving Consistent and Comparable CPU Evaluation}
\author{Chenxi Wang}
\affiliation{%
  \institution{Institute of Computing Technology, Chinese Academy of Sciences; State Key Lab of Processors, Institute of Computing Technology, Chinese Academy of Sciences; University of Chinese Academy of Sciences}
  \city{Beijing}
  \country{China}
}
\email{wangchenxi21s@ict.ac.cn}

\author{Lei Wang}
\affiliation{%
  \institution{The International Open Benchmark Council; Institute of Computing Technology, Chinese Academy of Sciences; University of Chinese Academy of Sciences}
  \city{Beijing}
  \country{China}
}
\email{wanglei_2011@ict.ac.cn}

\author{Wanling Gao}
\affiliation{%
  \institution{The International Open Benchmark Council; Institute of Computing Technology, Chinese Academy of Sciences; University of Chinese Academy of Sciences}
  \city{Beijing}
  \country{China}
}
\email{gaowanling@ict.ac.cn}

\author{Fanda Fan}
\affiliation{%
  \institution{The International Open Benchmark Council; Institute of Computing Technology, Chinese Academy of Sciences; University of Chinese Academy of Sciences}
  \city{Beijing}
  \country{China}
}
\email{fanfanda@ict.ac.cn}

\author{Yuchen Su}
\affiliation{%
  \institution{Institute of Computing Technology, Chinese Academy of Sciences;  University of Chinese Academy of Sciences}
  \city{Beijing}
  \country{China}
}
\email{suyuchen24s@ict.ac.cn}

\author{Yutong Zhou}
\affiliation{%
  \institution{Institute of Computing Technology, Chinese Academy of Sciences;  University of Chinese Academy of Sciences}
  \city{Beijing}
  \country{China}
}
\email{zhouyutong24s@ict.ac.cn}

\author{Yikang Yang}
\affiliation{%
  \institution{Institute of Computing Technology, Chinese Academy of Sciences;  University of Chinese Academy of Sciences}
  \city{Beijing}
  \country{China}
}
\email{yangyikang23s@ict.ac.cn}

\author{Jianfeng Zhan}
\authornote{Corresponding author.}
\affiliation{%
  \institution{The International Open Benchmark Council; Institute of Computing Technology, Chinese Academy of Sciences; University of Chinese Academy of Sciences}
  \city{Beijing}
  \country{China}
}
\email{zhanjianfeng@ict.ac.cn}




\begin{abstract}

\textcolor{black}{The challenge of CPU evaluation lies in the fact that user-perceived performance metrics can only be measured on an independently running system consisting of the CPU and other indispensable components, and hence it is difficult to accurately attribute the deviations in the evaluation outcomes to the differences between the CPUs. Our experiments reveal that the industry-standard CPU benchmark, SPEC CPU2017, suffers from a significant flaw: for the identical CPU, undefined configurations of other indispensable components introduce uncontrolled variability in evaluation outcomes (e.g., performance fluctuations ‌ranging from 4.01\% to 231.27\%), and the confounding from the other indispensable components significantly distorted the evaluation outcomes.}


We propose a rigorous CPU evaluation methodology.  Through theoretical analysis and pioneering controlled experiments, we systematically compare our methodology against four established methodologies: 
 the SPEC CPU 2017, two DOE variants, and one RCTs approach. The results show our methodology can achieve consistent and comparable evaluation outcomes, while others exhibit inherent limitations‌. 

\end{abstract}

\keywords{CPU Evaluation, SPEC CPU, Design of Experiments, Randomized Control Trials}

\maketitle

\section{Introduction}\label{sec:introduction}


\textcolor{black}{Rigorous evaluation methodologies like design of experiments (DOE)~\cite{telford2007brief} and Randomized Control Trials (RCTs)~\cite{stolberg2004randomized} have revolutionized fields ranging from pharmaceutical development to industrial engineering, while computer performance evaluation, particularly in CPU evaluation, remains predominantly based on empirical methodologies. As a widely recognized industry standard, the SPEC CPU2017 benchmark serves as a `standard athletic stadium' for CPU performance evaluation, utilizing 43 crafted workloads and 3 prescribed standard datasets (`test,' `train,' `ref'). }


\textcolor{black}{Unlike the drug evaluation, CPU evaluation presents unique challenges. In the field of drug evaluation, the evaluated object is the drug, while the affected object is the patient, independent from the evaluated object. In the field of CPU evaluation, the affected object is not straightforward, and stakeholders prefer user-perceived performance, e.g, the execution time of a typical workload, instead of the peak theory number, e.g., FLOPs, that can be directly reported on the CPU. Unfortunately, user-perceived performance metrics can only be measured on an independent running system consisting of the CPU and other indispensable components.}

\begin{figure}[h]
\centering
\includegraphics[scale=0.5]{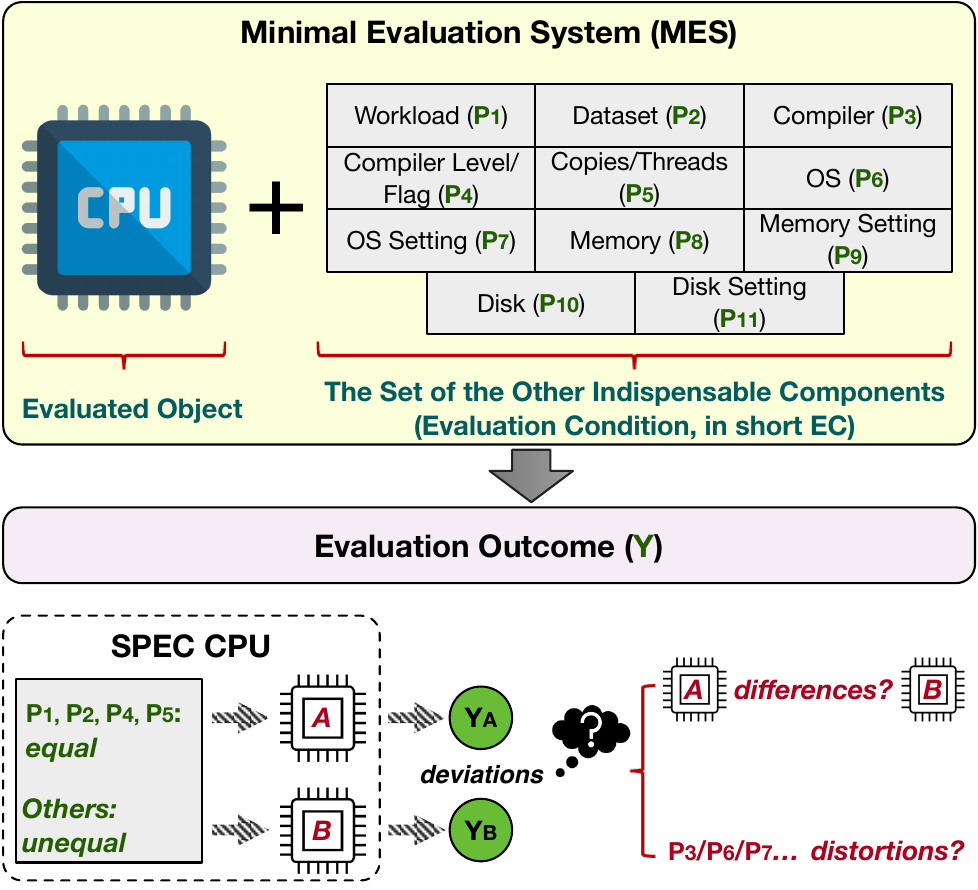}
\caption{
\textcolor{black}{The Challenge of CPU Evaluation lies in the fact that User-perceived performance metrics can only be measured on the \underline{m}inimal \underline{e}valuation
\underline{s}ystem (MES), which we formally define in Section~\ref{sec:introduction}, and hence it is difficult to accurately attribute the deviations in the evaluation outcomes on MES to the differences between the CPUs when the confounding from the other indispensable components of MES will distort the evaluation outcomes.}
}
\label{fig1-com}
\end{figure}

\textcolor{black}{In this article, we formally call the above independent running system  \textit{the \underline{m}inimal \underline{e}valuation \underline{s}ystem (in short, MES)}. An MES has three properties. First, it consists of the CPU and other indispensable components. Second, it can ensure independent running, and the user-perceived performance metric can be directly reported on that system. Third, it excludes the other components that have no impact on the user-perceived metrics.  }

\textcolor{black}{The challenge of the CPU evaluation is to accurately attribute the deviations in the evaluation outcomes on MES to the differences between the CPUs and avoid the confounding from the other indispensable components of MES. In the rest of this article, we call the set of the other indispensable components of MES \textit{the \underline{e}valuation \underline{c}ondition (in short, EC)}. In other words, when we apply the EC to the evaluated object, we construct an MES.   }
 

\textcolor{black}{Our experiments reveal that the industry-standard CPU benchmark, SPEC CPU2017, suffers from a significant flaw. Evaluating the identical CPU, the undefined EC configurations introduce uncontrolled variability in evaluation outcomes (e.g., performance fluctuations ‌ranging from 4.01\% to 231.27\%‌). That is to say, the SPEC CPU2017 methodology can not accurately attribute the deviations in the evaluation outcomes on MES to the differences between the evaluated objects, and the confounding from EC significantly distorted the evaluation outcomes.  
Those phenomena can also be observed from other widely-used benchmarks like PARSEC~\cite{bienia2008parsec}, CloudSuite~\cite{ferdman2012clearing}, MLPerf~\cite{reddi2020mlperf}, and BigDataBench~\cite{wang2014bigdatabench}.}



We propose a rigorous CPU evaluation methodology,  which we call CPU evaluatology, as it is inspired by the general evaluation framework named evaluatology~\cite{zhan2024evaluatology, zhan2024short,zhan2025fundamental}. \textcolor{black}{Different from the original evaluatlogy methodology (detailed in Section~\ref{evaluatology_difference}), our CPU evaluatology consists of four essential steps. The first step is to define the evaluated object. The purpose is to ensure
the instances to be evaluated belong to the same object and guarantee the apple-to-apple comparison. The second step is to define the MES with the two goals. First, it is to identify other indispensable components of MES  that can impact user-perceived performance metrics while excluding the irrelevant components. Second, it is to include diverse EC configurations that reflect the real-world distributions. The third step obtains the true MES evaluation outcomes in order to report user-perceived performance in real-world scenarios. We establish probabilistic bounds on the population mean \(\mu\) via the sample mean \(\bar{x}\) in order to ensure consistent true evaluation results. The fourth step is to attribute the deviation of the evaluation outcome to the differences between the evaluated objects. The purpose is to avoid the confounding of EC configurations and ensure the comparability of different evaluated objects.}

We also compare our CPU evaluatology against two DOE, and one RCTs methodologies through theory analysis and experiments.
The $2^kr$ factorial and general factorial designs and RCTs rely upon the evaluator's expertise. The former two leave the task of choosing the $k$ factor from $n$ factors to evaluators' domain-specific expertise. The latter may omit critical factors in the design of the evaluation configuration and hence introduce bias. The $2^kr$ factorial design is burdened by a rigid factor level partitioning without considering the heterogeneous characteristics of different factors, while the general factorial design is prohibitively costly as it mandates exhaustively traversing the entire factor configuration space. RCTs mandate random assignment of evaluation configuration to CPUs. In this context, comparing two CPUs across different evaluation configurations introduces systematic distortion due to unaccounted confounding variables.

Our experiments demonstrate that the $2^kr$ factorial design, while time-efficient, yields inadequate accuracy; \textcolor{black}{only 25.10\% of 99\% confidence intervals contain ground truth, that is, the mean of evaluation outcome under our constructed population of EC configurations}. Conversely, the general factorial design and the Randomized Controlled Trials (RCTs) achieve target accuracy but require exhaustive configuration space exploration, which becomes computationally prohibitive for large-scale configuration spaces such as our constructed billion-scale space. So, the dilemma of time costs and accuracy tradeoff makes DOE or RCTs to achieve consistent and comparable  CPU evaluation almost impossible due to prohibitive cost.

Our contributions are as follows.
\begin{enumerate}
\item \textcolor{black}{We present the unique challenge of the CPU evaluation} and formally define the CPU evaluation problem.
\item We propose a rigorous CPU evaluation methodology.
\item We reveal the limitations of four established evaluation methodologies through theory analysis in the CPU evaluation scenario.
\item For the first time, we compare SPEC CPU2017, two DOE, and one RCTs methodologies and reveal their limitations through controlled experiments in the CPU evaluation scenarios. 
\item We validate that our CPU evaluatology can achieve consistent and comparable CPU evaluation outcomes with acceptable time costs through controlled experiments.  
\end{enumerate}

The remainder of the paper is organized as follows. Section~\ref{Sec_P&M} formally defines the CPU evaluation problem. Section~\ref{Sec_M&R} discusses the motivation and related work. Section~\ref{CPU_Eva_Method} proposes our CPU evaluatology. Section~\ref{TheEvaluationsection} shows the evaluation and validation of our methodology against the other methodologies, and Section~\ref{Sec_Conclusion} concludes.



\section{Problem Definition}\label{Sec_P&M}

This section formally defines the CPU evaluation problem.




\textcolor{black}{We define the evaluation as the process of inferring the effect of the evaluated object on the affected object. In the drug evaluation, the evaluated object is the drug, while the affected object is the human body. The CPU evaluation differs significantly from that of drug evaluation. The evaluated object is the CPU; 
The peak theory metrics of a CPU, like FLOPS, can be directly reported on the CPU. However, the stakeholders prefer the user-perceived performance, e.g, the execution time of a typical workload. Unfortunately, user-perceived performance metrics can only be measured on the MES, as shown in Figure~\ref{fig1-com}. Only within the MES can we indirectly infer the CPU's specific contribution. So, the challenge of the CPU evaluation is to accurately attribute the deviations in the evaluation outcomes on the MES to the differences between the CPUs and avoid the confounding from the EC. }



\textcolor{black}{$\text{MES}_i = \mathrm{O}_i \times \text{EC}$, indicate the MES configuration space equals applying all EC configurations in $\text{EC}$ to $\mathrm{O}_i$. Specifically, $\text{MES}_i = \mathrm{O}_i \times P_1 \times P_2 \times \cdots \times P_n$ represent a well-defined MES configuration space. }

\textcolor{black}{$\text{O}_i = \{ \mathrm{o}_{i1}, \mathrm{o}_{i2}, \ldots, \mathrm{o}_{ik}, \ldots \} $ represents a well-defined evaluated object configuration space, where i refers to object selection and k enumerates their possible configurations. For example, turbo boost mode and default operational mode of the same CPU  are treated as two separate configurations in the $\mathrm{O}_i$ configuration space. In practice, the selection of the evaluated object could be the sole CPU or several components in addition to the CPU, as specified in the SPEC methodology, that are combined to pursue the best performance. }

\textcolor{black}{$P_j = \{ \mathrm{p}_{j1}, \mathrm{p}_{j2}, \ldots, \mathrm{p}_{jk}, \ldots \}$, where $j\in\{1, 2, 3, ..., n\}$, is the $j^{\text{th}}$ indispensable component of MES,  or the $j^{\text{th}}$ element of EC, where k enumerates their possible configurations. The EC configuration space is $\text{EC} = P_1 \times P_2 \times \cdots \times P_n$. We have $\text{EC} = \{\text{ec}_{i}\}$, where $i\in\{1,2,3,...,m\}$. There are a total of $m=\prod_{j=1}^{n} m_j$ configurations $ec_{i}$ in $\text{EC}$, $m_j$ is the configuration size of each component $P_j$.}

\textcolor{black}{The evaluation outcome deviation between two evaluation objects, ${o}_{ij}$ and ${o}_{ik}$, is defined as the difference in their user-perceived performance metrics when applying each equivalent EC configuration to the two CPUs in one pair of experiments. }





\section{What is Wrong with the existing work?}\label{Sec_M&R}

\subsection{The Flaw of SPEC CPU methodology}\label{Sec-Mot}

Lacking the rigor methodology, computer performance evaluation relies on empirical methodologies, which are often called benchmarks~\cite{panda2018wait,ferdman2012clearing,bienia2008parsec,bienia2008parsec1,woo1995splash,hoste2006comparing,hoste2007analyzing,hoste2006performance,farhaditrace,hoste2007microarchitecture,shao2013isa,phansalkar2005measuring,phansalkar2007analysis,limaye2018workload,wang2023wpc,campanoni2010highly}. The SPEC CPU is the most prominent CPU benchmark. Over time, six versions of the SPEC CPU benchmark have been released, including SPEC CPU89~\cite{SPECCPU89}, SPEC CPU92~\cite{SPECCPU92}, SPEC CPU95~\cite{SPECCPU95}, SPEC CPU2000~\cite{SPECCPU2000}, SPEC CPU2006~\cite{SPECCPU2006}, and SPEC CPU2017~\cite{SPECCPU2017}.

SPEC CPU2017 categorizes its benchmark suites based on calculation type into integer (int) and floating-point (fp) workloads. According to the evaluation metrics, SPEC further divides the workloads into the SPECrate sub-suite (single-thread, multi-copies workloads; higher scores mean that more work is done per unit of time) and the SPECspeed sub-suite (single-copy, multi-thread workloads; higher scores mean that less time is needed). Therefore, SPEC CPU2017 comprises four sub-suites: intrate, fprate, intspeed, and fpspeed~\cite{SPECCPU2017}. 
The standard output files of SPEC CPU2017 will report the composite score (run each workload three times and use the median score; the sub-suite's score is the geometric mean of all its workloads' scores) of these four sub-suites.

Instead of choosing the sole CPU as the evaluated object, SPEC CPU takes several components together as the evaluated object, consisting of the workload ($P_1$), dataset ($P_2$), compiler levels/flags ($P_4$), copies/threads  ($P_5$), and the CPU itself. SPEC CPU2017 provides the recommended configuration for $P_1$, $P_2$, $P_4$, $P_5$ while keeping the EC configuration undefined. In detail, it provides the specified workload, the corresponding `ref' datasets, `-O3' compiler levels/flags, and sets the number of copies/threads to the maximum number of hardware threads supported by the CPU.

\textcolor{black}{Our experiments reveal that following the SPEC CPU methodology, the evaluation outcomes of the identical CPU exhibit substantial variability. The evaluated target is an x86 CPU (CPU A), whose machine information is shown in Table~\ref{Machine-Inf} in Section~\ref{TheEvaluationsection}.}



\begin{figure}[h]
\centering
\includegraphics[scale=0.42]{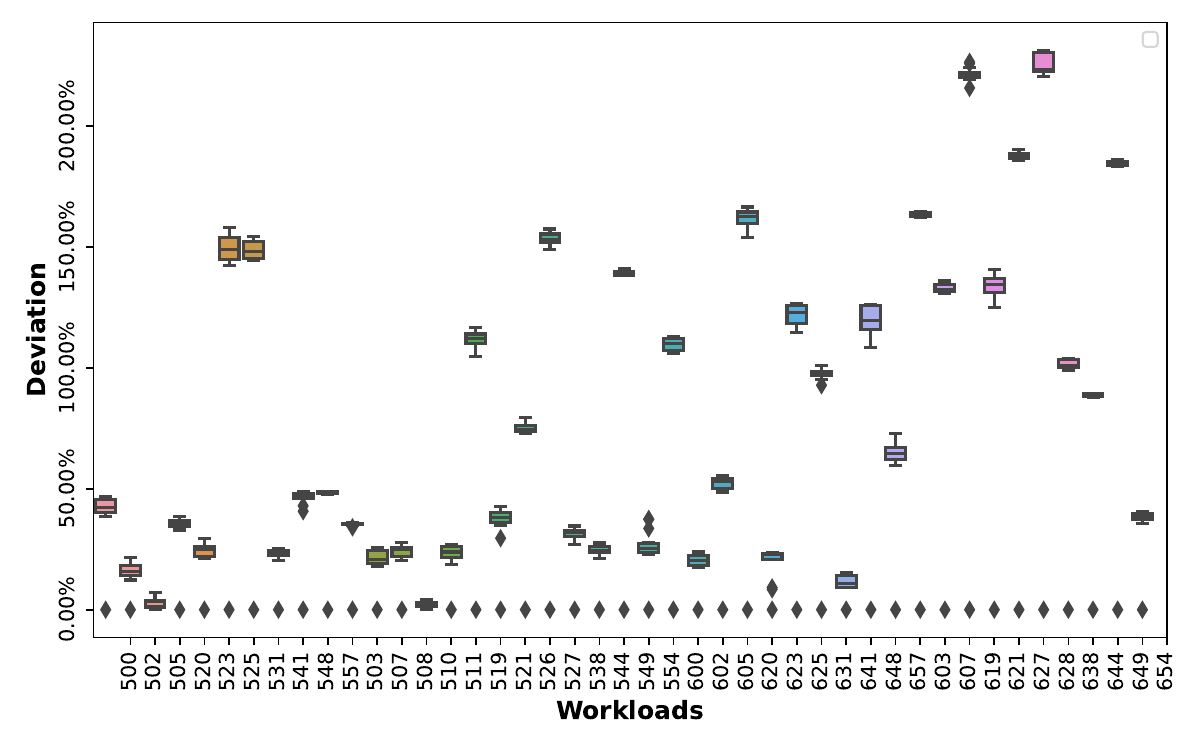}
\caption{
\textcolor{black}{In this setting, the evaluated object consists of CPU A, SPEC CPU2017 workloads, `ref' dataset, `-O3' compiler level/flag,  and 56 copies/threads strictly adhering to the SPEC CPU2017 methodology. We found that the user-perceived performance significantly differs due to uncontrolled EC configurations, which confounds the evaluation outcomes. We collect the experiments from our experiments and the official release of SPEC CPU2017~\cite{CPU2017_results}. We show the box plot~\cite{williamson1989box} of the deviation. 
}
}
\label{mot-dif}
\end{figure}

\textcolor{black}{For the identical evaluated object, we observe that the wildly varied user-perceived performance fluctuations are due to the 16 different EC configurations within the MES}.

\textcolor{black}{As shown in Figure~\ref{mot-dif}, across all SPEC CPU2017 workloads, evaluation outcomes for the identical evaluated object under a single workload varied by 4.01\% to 231.27\%. Notably, this variation occurred while keeping the identical evaluated object, with outcomes peaking at 231.27\% variation, which is significantly fluctuating. Moreover, the final SPEC report demonstrated notable performance discrepancies across sub-suites: 43.22\% variance in SPECfprate, 58.01\% deviation in SPECintrate, 71.63\% difference in SPECfpspeed, and a substantial 141.66\% variation in SPECintspeed.}

\textcolor{black}{From this experiment, we have two observations. First, even if the evaluated object includes the workload ($P_1$), dataset ($P_2$), compiler levels/flags ($P_4$), copies/threads ($P_5$), and the CPU itself, which are often implemented to achieve the best CPU performance for different vendor, the variability of other component configuration will significantly affect the evaluation outcome of the same evaluated object. Second, without the components of EC, we cannot obtain the evaluation outcome of the evaluated objects. So diverse EC configurations that reflect the real-world distribution should be included in the evaluation.}





\subsection{The Pitfalls of DOE and RCTs}

Rigorous evaluation methodologies, such as the design of experiments (DOE)~\cite{telford2007brief} and Randomized Control Trials (RCTs)~\cite{stolberg2004randomized}, are widely used in many domains. 

Canonical DOE methodologies encompass $2^ 
kr$ factorial design and general factorial design~\cite{telford2007brief,jain1991art}. The former involves selecting k factors, where each factor's levels are divided into high and low intervals. For each experimental run, one level is sampled from each factor's high and low intervals, generating $2^k$ configurations through a combinatorial pairing of the $k$ factors. Each configuration undergoes $r$ experimental repetitions. 

The general factorial design maintains $k$ factor selection but requires each factor to traverse all its levels, thereby constructing a complete configuration space of all possible factor level combinations. Experiments are conducted within this entire configuration space, with each configuration also repeated $r$ times.

Similarly, RCTs employ random sampling without replacement to select a specified number of configurations 
from the entire space. These configurations are then randomly divided into control and treatment groups, with experiments conducted under both groups' configurations, each repeated $r$ times~\cite{stolberg2004randomized}. 

\subsubsection{The limitations in the $2^kr$ factorial design}

For $2^kr$ factorial design, in building configuration space, it selects $k$ factors  $P_i$, where $i \in\{1,\ldots, k\}$, and $k \leq n$. 

For each selected factor $P_i$, it divides $P_i$ into a low-level and a high-level. For each $P_i$, randomly select a configuration from both the low-level and high-level. Thus,
for each $p_{ij}$, where $i \in \{1,\ldots, k\},$ $j \in \{low, high\}$. Finally, the evaluator applies all $2^k$ configurations to both $o_{11}$(CPU 1) and $o_{12}$(CPU 2), repeating the experiment of each configuration $r$ times. 

There are two limitations. First, choosing the $k$ factor from $n$ factors depends on the evaluator's expertise. Increasing the number of factors k causes an exponential explosion of the configuration space, leading to prohibitive cost escalation, while reducing k risks comparative invalidity due to uncontrolled non-factorial component variations between CPU evaluations. Second, there is a rigid Factor Level Partitioning without considering the heterogeneous characteristics of different factors. It divides each factor's configurations into high/low levels, whether through uniform distribution or expert-driven differential allocation, significantly impacting evaluation outcomes reliability.

\subsubsection{The limitations in the general factorial design}

For the general factorial design configuration space, it
selects $k$ factors $P_i$, 
where $i \in\{1,\ldots, k\}$ and $k \leq n$.
Each factor $P_i$ has $m_i$ factor-levels, using all factor-levels of $P_i$, $k$ factors construct a total of $m=\prod_{\substack{i=1}}^k m_i$ configurations, Finally, the evaluator applies all $m$ configurations to both $o_{11}$(CPU 1) and $o_{12}$(CPU 2), repeating the experiment of each configuration $r$ times. 

The first limitation of the $2^kr$ factorial design still applies to the general factorial design.  Second, it is prohibitively costly as it mandates exhaustive traversal the entire factor configuration space, which renders practical implementation prohibitively costly or even impractical.

\subsubsection{The limitations in the RCTs design} 

Different from the drug evaluation, in which patients (units) receive different drug interventions, in the CPU evaluation scenario, EC configurations, serving as the units, are assigned to different CPUs as the control and treatment interventions.

RCT do not consider the configuration space that constitutes a minimum evaluation system. Instead, it randomly divides the configuration space $C$, which is built by the evaluators relying upon their expertise,  into control $C_{control}$ and treatment $C_{treatment}$ groups using without-replacement sampling, therefore, $C=C_{control} \cup C_{treatment}$.  Finally, the evaluator applies all evaluation configurations $c_{control}$ to CPU 1, where $c_{control} \in C_{control}$, and applies all configurations $c_{treatment}$ to CPU 2, where $c_{treatment} \in C_{treatment}$. 

RCTs have two limitations. First, The design of its configuration space relies upon the evaluator's expertise, and it may omit critical factors and hence introduce bias. Second, RCTs mandate random assignment of evaluation configuration to CPUs. Comparing two CPUs across different evaluation configurations introduces systematic distortion due to unaccounted confounding variables.

\subsection{Other related works}

Wang et al.~\cite{wang2022study} and Benson et al.~\cite{benson2024surprise} highlighted a significant issue in database performance evaluation: the experimental configuration has a substantial impact on evaluation outcomes. They discussed potential methods to address this problem. However, they did not provide a systematic methodology for achieving consistent and comparable evaluation outcomes. Hennessy et al.~\cite{hennessy2019new} changed the algorithm and implementation of matrix multiplication workload and observed several orders of magnitude performance gaps, which motivated our comprehensive consideration of all valid configurations when designing the population of EC configurations. 

Chen et al.~\cite{chen2014statistical} proposed a novel non-parametric statistical framework for comparing computer performance with small sample sizes, which provided us with ideas for our sampling method. Thomas et al.~\cite{diciccio1996bootstrap} emphasized that the bootstrap method can automatically construct highly accurate approximate confidence intervals in complex data structures and probabilistic models, which provided a reference for calculating confidence intervals.
Mytkowicz et al.~\cite{mytkowicz2009producing} found the measurement bias under different experimental setups, and proposed methods to avoid and detect measurement bias. Their research pointed out the influence of confounding factors on evaluation outcomes, promoting our thinking on well-defining the EC configuration space.

\section{\textcolor{black}{The CPU evaluatology}}\label{CPU_Eva_Method}

This section presents our innovative CPU evaluation methodology, which we call CPU evaluatology as it is inspired by the evaluatology~\cite{zhan2024evaluatology}; however, it has significant differences as discussed in Section~\ref{evaluatology_difference}.


\subsection{How to address the CPU evaluation problem?}\label{basic_idea}

\textcolor{black}{Considering two CPUs $C\_a$  and $C\_b$, from two CPU vendors a and b, the goal of CPU evaluation is to attribute the deviations in evaluation outcomes observed on the MES to the differences between $C\_a$  and $C\_b$, and avoid the confounding from EC that distorts the evaluation outcomes.
}


\textcolor{black}{The most straightforward solution is to specify the sole CPU, $C\_a$  and $C\_b$, as the evaluated object while keeping each EC configuration the same, so we can accurately attribute the deviations in the evaluation outcomes to the differences between $C\_a$  and $C\_b$. However, there is a huge configuration space for both CPU and EC, and hence, the evaluation cost is unaffordable. }

\textcolor{black}{Someone may argue it is not fair to only include the sole CPU as the evaluated object, because some components of EC are closely related to the CPU design. In practice, different CPUs have different design capabilities for multi-threading. For example, CPU $C\_a$ has more hardware threads than that of $C\_b$. If we mandated the copies/threads to be the same in all EC configurations, CPU $C\_a$ can not achieve its peak raw performance in some cases. A natural solution is to combine several components together as the evaluated object, so we can compare the best performance of each CPU. For example, the SPEC CPU methodology includes the copies/threads with the CPU as the evaluated object. However,  a CPU with the combined components can not always achieve the best performance in all EC configurations, as we showed in Table~\ref{Best-Per_Thread}. \textit{The CPU behaves well in some cases while degrading in other cases.} The diverse configurations of EC are still essential, and one EC configuration can not represent all. }

\begin{table}[]
\footnotesize
\caption{\textcolor{black}{The thread configurations when the CPU achieves best performance in 621.wrf\_s workload's different datasets.}}
\begin{tabular}{cccc}
\hline
\multicolumn{2}{c}{CPU A}                                                       & \multicolumn{2}{c}{CPU B}                                                       \\ \hline
\multicolumn{1}{l}{Dataset} & \multicolumn{1}{l}{Best Performance Thread} & \multicolumn{1}{l}{Dataset} & \multicolumn{1}{l}{Best Performance Thread} \\ \hline
1                                 & 55                                          & 1                                 & 62                                          \\
2                                 & 56                                          & 2                                 & 62                                          \\
3                                 & 56                                          & 3                                 & 61                                          \\
4                                 & 56                                          & 4                                 & 61                                          \\
5                                 & 53                                          & 5                                 & 61                                          \\
6                                 & 35                                          & 6                                 & 62                                          \\
7                                 & 56                                          & 7                                 & 61                                          \\
8                                 & 56                                          & 8                                 & 61                                          \\
9                                 & 55                                          & 9                                 & 61                                          \\
10                                & 54                                          & 10                                & 61                                          \\ \hline
\end{tabular}
\label{Best-Per_Thread}
\end{table}



\textcolor{black}{In the next subsection, we will present our solution.  The focus of this article is to emphasize the rigorous methodology design and its practical implications. We plan to release a new open-source CPU benchmark suite. Currently, we reuse the SPEC CPU2017 benchmark suite.}

\textcolor{black}{Our methodology has three key considerations. First, to attribute the deviation in the evaluation outcomes observed on the MES to the differences between the evaluated objects, we mandate each equivalent EC configuration to be applied to the evaluated objects. Second, since one specific EC configuration can not represent all, we incorporate diverse EC configurations that reflect the real-world distributions. Third, we balance the evaluation accuracy with the evaluation cost.}

\subsection{The four essential steps of CPU evaluatology}\label{CPU_eva_four_steps}

\begin{figure*}[h]
\centering
\includegraphics[scale=0.41]{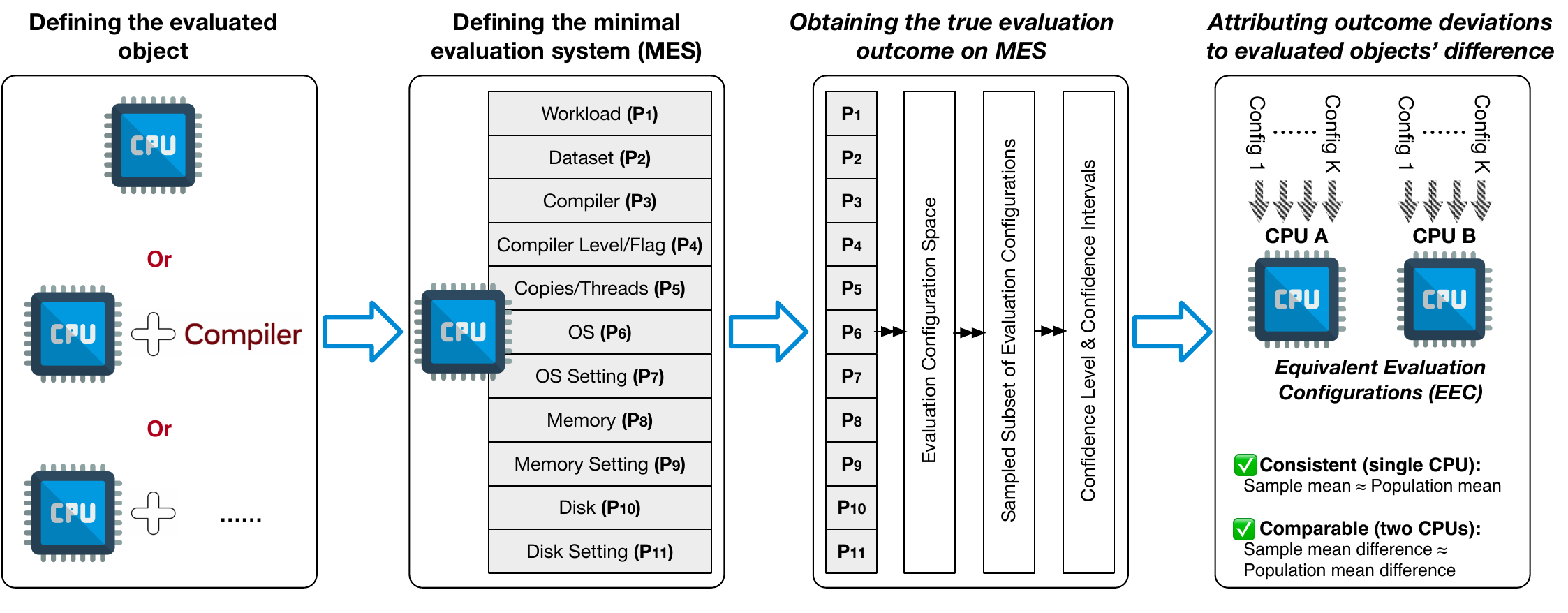}
\caption{\textcolor{black}{The four essential steps of CPU Evaluatology.}
}
\label{cpu-evaluatology}
\end{figure*}

\textcolor{black}{As shown in Figure~\ref{cpu-evaluatology}, CPU evaluatology consists of four essential steps. When explaining the general procedure, we also demonstrate how to use this methodology. }

\subsubsection{Defining the evaluated object}\label{step_one}

\textcolor{black}{It is essential to characterize and define what the evaluated object is. To achieve that goal, we need to identify the primary components of the evaluated object and how they work together. Without the clear definition of an evaluated object, we cannot ensure different instances belong to the same object and hence cannot guarantee an apple-to-apple comparison. }

\textcolor{black}{For the CPU evaluation, even if the evaluated object only includes the CPU, it still includes many configurations, such as turbo boost, thermal throttling, or power capping. 
Let $\text{O}_i = \{ \mathrm{o}_{i1}, \mathrm{o}_{i2}, \ldots, \mathrm{o}_{ik}, \ldots \} $ represents a well-defined evaluated object configuration space, where i refers to object selection and k enumerates their possible configurations. For example, turbo boost mode and default operational mode of the same CPU  are treated as two separate configurations in the CPU configuration space. }

\textcolor{black}{In practice, to reduce the evaluated object configuration space, a specific configuration of the evaluated object that achieves the best performance in some EC configurations is adopted while hiding many other evaluated object configurations.  However, this choice has a pitfall because a specific evaluated object configuration can not achieve the best performance in all EC configurations. }



\subsubsection{Defining MES}\label{step_two}

\textcolor{black}{In Section~\ref{sec:introduction}, we have specified three properties of the MES. The essence of this step is to remove irrelevant components from the MES so we can reduce the cardinality of the EC configuration space and hence reduce the evaluation cost. The MES consists of the evaluated object $O_i$ and the EC. }

\textcolor{black}{Another decision point is how to partition the MES into the evaluated object and the EC. In practice, the industry tends to combine several components in addition to the CPU together as the evaluated object, so as to pursue the best performance of the CPU. As specified in the SPEC methodology, the evaluated objects include the CPU in addition to the workload ($P_1$), dataset ($P_2$), compiler levels/flags ($P_4$), copies/threads ($P_5$).   However, this choice has a side effect: it cannot distinguish the responsibility or attribute contributions of different components in the evaluation outcome. }


\textcolor{black}{We exclude the networks from the MES because the workloads only need to be executed locally when evaluating a CPU. Finally, we decide the MES cardinality $n=11$ when the evaluated object $O_i$ includes a sole CPU. EC includes workload ($P_1$), dataset ($P_2$), compiler ($P_3$), compiler levels/flags ($P_4$), the number of copies/threads ($P_5$),  OS ($P_6$), OS setting ($P_7$), memory ($P_8$), memory setting ($P_9$), disk ($P_{10}$), and disk setting ($P_{11}$).}


\textcolor{black}{For each component, we further determine each component's configuration size $m_i$ to systematically construct the valid EC configuration space. In practice, the valid configuration space encompasses diverse scenarios that reflect the real-world user requirement distributions. In practice, to reduce the evaluation cost, we could only choose the TOP N  configurations according to the cumulative distribution function (CDF).}

For the workload ($P_1$), this article does not intend to design the CPU benchmark from scratch. Instead, we retained 43 SPEC CPU2017 workloads. For the dataset ($P_2$), as the realistic applications have varying sizes of data inputs instead of fixed ones, we implemented a dataset generator for all SPEC CPU2017 workloads, and the size ranged from the minimum permissible size per workload to a maximum of 5 times the SPEC CPU2017 `ref' dataset size. This wide dataset range is large enough to reflect real-world user requirement distributions. The compiler level/flag ($P_4$) was shifted from `-O1', `-O2', and `-O3' as we found many workloads achieved better performance in different settings. For example, we found that the 648.exchange2\_s achieved better performance in `-O1', and 649.fotonik3d\_s achieved better performance in `-O2' than `-O3' on CPU A. The number of copies/threads ($P_5$) spanned 131 discrete values: integers 1 to 128 supplemented by 200, 256, and 300, determined by mainstream server CPU core counts~\cite{aa} and our machine's memory constraints. \textcolor{black}{We evaluate two CPUs with the default settings, and any other CPUs or CPU settings could be evaluated in the same way, without losing the generality of our methodology.}

\textcolor{black}{Meanwhile, to simplify the EC configuration space, we chose to make the compiler ($P_3$), OS ($P_6$), OS setting ($P_7$), memory ($P_8$), memory setting ($P_9$), disk ($P_{10}$), and disk setting ($P_{11}$) to a specific setting. This choice does not intend to exclude the other settings. Instead, we avoid the explosion of the configuration space without losing the generality of the methodology.}

\textcolor{black}{We treat the outcome under each EC configuration $ec_{i}$ as a valid outcome. Only reporting the CPU performance under a specific configuration, e.g., the recommended configuration in SPEC CPU2017,  will hide many scenarios in which the CPU or the difference between two CPUs may behave significantly differently.  We cannot discard any one of the evaluation outcomes because each one reflects the CPU performance under a different scenario. }

\textcolor{black}{Based on the above decision, we constructed a population of billion-scale valid EC configurations.}


\subsubsection{Obtaining the true evaluation outcome on MES}\label{step_three}
\textcolor{black}{}

\textcolor{black}{The evaluation outcomes can only be obtained under a specific MES configuration. Each configuration within the MES configuration space yields valid outcomes that reflect the distribution of real-world user scenarios. Consequently, the evaluation outcomes derived from each MES configuration constitute the true evaluation outcome of the MES. }

\textcolor{black}{In practical implementations, we could adopt a representative configuration on the evaluated object configuration space $O_i$ (e.g., CPU) to reduce the MES configuration space and hence reduce the evaluation cost. For example, we could select the most commonly used evaluated object configuration that achieves top 20\% performance across over 80\% of scenarios based on the cumulative distribution function (CDF) - and maintain this configuration consistently throughout experiments.}

\textcolor{black}{After determining the representative evaluated object configuration $O_i$ (e.g., CPU), we consider each configuration $ec_{i}$ within the EC configuration space $EC$ valid. The evaluation outcome of a CPU is an aggregate of the evaluation outcome under each EC configuration $ec_{i}\in EC$, which constitutes the true evaluation outcome of a CPU. Since the space of $EC$ is huge, we need to find a representative configuration sample to infer the true evaluation outcome of the configuration population.}

\textcolor{black}{We utilize confidence level and confidence intervals to report the evaluation outcomes under the subset of EC configurations. The aim is to establish probabilistic bounds for a population mean estimation through the sample mean. We calculate the arithmetic mean as the sample mean for users who are concerned about the average performance. For users with other concerns, you may choose another policy, e.g., a weighted mean based on the weight preferred by the user.}


When a well-defined EC configuration is applied to a specific CPU, the evaluation outcome can be reported by taking the mean or median of multiple experiments~\cite{telford2007brief}. In CPU evaluation, when a well-defined EC configuration is applied to a specific CPU, it yields a true value. However, due to the complexity of the OS, software stack, and the CPU's mechanisms, such as superscalar and out-of-order execution, there is inherent randomness in the instruction stream executed by the CPU pipeline. This randomness causes the evaluation outcomes to fluctuate slightly around the true value. 

To minimize this fluctuation and report the evaluation outcome for a specific CPU under a well-defined EC configuration, multiple experiments should be conducted, and their mean or median should be taken. Using the mean as the outcome can balance out the random noise from different experimental runs, but may be affected by outliers and does not represent an actual experimental result. Using the median as the outcome is not influenced by outliers and is an actual experimental result, but it may not utilize all data points, which could impact balancing randomness. In practice, the choice between these methods depends on the specific circumstances. For example, performance monitoring tools like perf~\cite{de2010new} report means, while SPEC CPU2017 uses the median of scores from individual workloads to calculate the final score~\cite{SPECCPU2017}.

When a subset of EC configurations is applied to a specific CPU, the evaluation outcomes can be determined using confidence intervals at a given confidence level, for example, 95\%. In CPU evaluation, using only one specific EC configuration is unrepresentative and not consistent, and using all valid EC configurations is impractical. Therefore, a subset of EC configurations is sampled for the evaluation. According to our CPU evaluatology, each well-defined EC configuration yields a valid evaluation outcome. We further calculate the confidence intervals of these outcomes at a given confidence level to represent the evaluation outcomes for the subset of EC configurations. 
 
All values have the same probability of being significant at the given confidence level. Once all valid EC configurations are well-defined, the population mean is a constant. Under the Central Limit Theorem~\cite{jain1991art}, repeated sampling from this population yields sample means of large samples that follow an approximately normal distribution. Confidence intervals derived from these sample means provide probabilistic bounds for estimating the population mean.

Due to the extremely high costs of exploring all valid EC configurations, it is essential to propose a sampling approach to reduce the entire EC configuration space to an affordable subset of configurations. There are many sampling approaches that we can utilize in statistics textbooks~\cite{moore2007basic}. 

We implemented iterative stratified random sampling~\cite{moore2007basic} to select the subset of EC configurations. As discussed in the first step, our methodology preserves 43 SPEC CPU2017 workloads while defining the parameter range for the dataset, compiler levels/flags, and the number of copies/threads. We also make other system configuration components (the compiler, OS, OS settings, memory, memory settings, disk, and disk settings) the same. 

To ensure comparability with SPEC CPU2017 outcomes and address workload heterogeneity, we performed stratified sampling with each SPEC workload as a stratum. Specifically, each sampling iteration randomly selected one value from each varied component (dataset, compiler levels/flags, copies/threads) per workload, generating a total of 43 configurations per iteration. Through 32 iterations, we obtained more than 30 samples per workload (a total of 1,376 configurations), satisfying large-sample criteria for statistical significance.

\subsubsection{Attributing the deviation of the evaluation outcome to the differences between the evaluated objects}\label{step_four}

\textcolor{black}{To accurately attribute the deviation of the evaluation outcome to the differences between the evaluated objects, we must ensure each equivalent EC configuration is applied to two evaluated objects in a pair of experiments. The purpose is to eliminate the confounding from EC configurations in attributing the deviation of the evaluation outcome to the differences between the evaluated objects. }

We employed execution time differences rather than execution time ratios for CPU comparisons in a pair of experiments. We present why we do not use ratio-based comparisons between two CPUs. We compare two cases using CPU A as the denominator (baseline) or using CPU B as the denominator.

As discussed in Section~\ref{Sec-Mot}, SPEC CPU2017 divides its 43 workloads into four sub-suites (SPECintrate, SPECfprate, SPECintspeed, and SPECfpspeed), and the final evaluation outcomes are based on these sub-suites. Similarly, we categorized the 1,419 configurations in our experiment according to the SPEC CPU2017 sub-suites, obtaining the distribution of execution time difference or execution time ratio between CPUs A and B for each sub-suite, along with their 95\% confidence intervals. 

When using CPU A as the denominator (baseline), ratio-based 95\% confidence intervals are: SPECintrate: (0.793, 0.881), SPECfprate: (1.054, 1.132), SPECintspeed: (0.750, 0.799), SPECfpspeed: (1.064, 1.120). When using CPU B as the denominator (baseline), intervals become: SPECintrate: (1.413, 1.591), SPECfprate: (1.006, 1.076), SPECintspeed: (1.383, 1.518), SPECfpspeed: (0.946, 1.012). This demonstrates ratio asymmetry. 1) Intervals are not reciprocal despite reciprocal per-configuration ratios. 2) Distribution histograms and curves exhibit morphological changes when alternating the baseline CPU. 3) Contradictory conclusions emerge.

For the SPECfprate sub-suite, when choosing CPU A as a baseline, confidence intervals greater than 1 imply CPU B is slower. However, when choosing CPU B as a baseline, confidence intervals greater than 1 imply that CPU A is slower. Moreover, for the SPECfpspeed sub-suite, when choosing CPU A as a baseline, confidence intervals greater than 1 imply CPU B is slower. When choosing CPU B as a baseline, confidence intervals including 1 imply that there is no significant performance difference between CPU A and CPU B. Conversely, difference-based analysis preserves mathematical symmetry. Alternating baseline CPU yields exact negative values for both per-configuration differences and confidence intervals, ensuring consistent conclusions across all SPEC CPU2017 sub-suites. Moreover, ratio-based analysis remains susceptible to Simpson's paradox~\cite{pearl2018book}, where CPU A demonstrates superior performance per individual workload, yet CPU B exhibits better aggregate performance across all workloads.

When we use difference-based comparisons between two CPUs, the above abnormal behavior can not be observed. 


\textcolor{black}{When evaluated under equivalent EC configurations. In a pair of experiments, for the equivalent EC configuration, evaluation outcome differences between the two CPUs are calculated. The controlled environment ensures statistically non-confounding comparisons between two CPUs. Under different EC configurations, these evaluation outcomes collectively form a performance sample, from which we derive confidence intervals at specified confidence levels. The intervals are compared against zero to determine performance relationships. We have two guidelines as follows.}

\textcolor{black}{First,  \textit{the intervals include 0}. It indicates no significant performance difference exists between CPUs within the subset of EC configurations, also suggesting no significant performance difference exists between CPUs under the population of EC configurations.}

\textcolor{black}{Second,  \textit{the intervals exclude 0}. It indicates that a significant performance difference is detected. We take execution time as an example. For negative intervals (left of 0), the Minuend CPU outperforms. For positive intervals (right of 0), the Subtrahend CPU (comparator) outperforms. }

\textcolor{black}{Please note that there are two constraints in our CPU evaluatology. First, we must ensure comparisons of two CPUs under equivalent EC configuration in a pair of experiments to gain a comparison sample and distribution. 2) Evaluation outcome differences (not ratios) enable symmetric statistical analysis. Arithmetic mean (confidence intervals) of differences preserve sign reversal when swapping CPU roles, whereas ratios lack this symmetry due to denominator variability across EC configurations. This asymmetry could yield conflicting conclusions when alternating the baseline CPU.}

\subsection{The difference of CPU evaluatology from the generalized evaluatology}\label{evaluatology_difference}

Zhan et al.~\cite{zhan2024evaluatology, zhan2024short,zhan2025fundamental} define evaluation as “conducting deliberate experiments where well-defined evaluation (condition) configurations are applied to a well-defined object," and propose a generalized evaluation framework: Evaluatology. 
Our work is based upon evaluatology. However, our work has significant differences as follows. 

\textcolor{black}{First, we distinguish the differences between the drug evaluation and CPU evaluation. We reveal that the challenge of CPU evaluation lies in the fact that user-perceived performance metrics can only be measured on the MES, and hence, it is difficult to accurately attribute the deviations in the evaluation outcomes to the differences between the CPUs and avoid the confounding from the EC configurations.}

\textcolor{black}{Second, the essence of our methodology has significant differences. Evaluatology provides a generalized framework for evaluation methodologies, without distinguishing between different evaluated objects, such as a drug or a CPU. Its methodology comprises four key steps. The first step establishes the EC. The second step constructs the equivalent EC. The third step eliminates all confounding variables. The fourth step defines value functions that capture the primary concerns.}

\textcolor{black}{Instead, CPU evaluatology redefines the CPU evaluation problem by establishing paradigms to address it. Its methodology comprises four key steps. The first step is to define the evaluated object. The second step is to define the MES. The third step obtains the true evaluation outcomes on MES. The fourth step is to attribute the deviation of the evaluation outcome to the differences between the evaluated objects. In this manner, CPU evaluatology formally defines the CPU evaluation problem and proposes a rigorous methodology for its assessment.}

\subsection{Discussions on Real-world Application of CPU evaluatology}\label{evaluatology_application}

\textcolor{black}{This article emphasizes the rigorous methodology in the CPU evaluation. We are planning to release an open-source CPU benchmark suite with the following considerations shortly.} 

\textcolor{black}{First, we will define several categories of the evaluated object. For example, we consider the sole CPU as a category, or a group of components, including the CPU, compiler, as another category of the evaluation object. }

\textcolor{black}{Second, for each category of evaluated object, we will provide a sample of well-defined and well-documented EC configurations that reflect the real-world distributions. }

\textcolor{black}{Third, for each CPU vendor or researcher, when they submit the evaluation outcome, they must choose the category of the evaluated object and apply the same sample of EC configurations to their evaluated objects. Finally, they must submit all evaluation outcomes under the prescribed sample of EC configurations. }

\textcolor{black}{Fourth, to reduce the evaluation cost, the vendor or researcher could only apply the TOP N EC configurations that reflect the real-world distributions to the evaluated object. However, the category of the evaluated object and the EC configuration must be well-defined and well-documented, so we can ensure the evaluation outcomes are comparable. }

\section{Evaluation}~\label{TheEvaluationsection}

In this section, we evaluate our proposed CPU evaluatology. Specifically, we aim to address the following two questions: 1) Can our CPU evaluatology deliver consistent and
comparable evaluation outcomes? 2) How does CPU evaluatology perform in comparison to four established methodologies: SPEC CPU2017, two DOE, and one RCTs methodologies—in terms of delivering consistent and comparable evaluation outcomes?

To answer these questions, we evaluate two CPUs with distinct architectural: CPU A (x86-based, 2 sockets × 14 cores/socket × 2 threads/core) and CPU B (ARM-based, 2 sockets × 32 cores/socket × 1 thread/core), \textcolor{black}{details in Table~\ref{Machine-Inf}}. 
\textcolor{black}{The configurations of the other components like memory and disk hold the same.}
All experiments are based on the valid configurations constructed via the four-step CPU evaluatology procedure described in Section~\ref{CPU_eva_four_steps}. 

\begin{table}[]
\centering
\caption{\textcolor{black}{Machine information in our experiments.}}
\begin{tabular}{lll}
\hline
\textbf{Machine}  & \textbf{CPU A}            & \textbf{CPU B}         \\ \hline
Architecture                &   x86       & ARM                \\
Socket(s)          & 2                             & 2                          \\
Core(s) Per Socket & 14                            & 32                         \\
Thread(s) Per Core & 2                             & 1                          \\
L1 (Per Core)                 & 32KiB I, 32KiB D              & 64KiB I, 64KiB D           \\
L2 (Per Core)                & 1MiB                          & 512KiB                     \\
L3 (Per Socket)                 & 19.25MiB Unified  & 32MiB Unified  \\
Memory             & 384GiB                        & 384GiB                     \\
Disk               & 2TiB                          & 2TiB                       \\
Compiler           & GCC9                     & GCC9                  \\ 
\hline
\end{tabular}
\label{Machine-Inf}
\end{table}

\subsection{Can CPU evaluatology and SPEC CPU2017 achieve consistent and comparable evaluation outcomes?}~\label{Mean_&_Confidence}

The purpose of this experiment is to demonstrate that CPU evaluatology can achieve consistent and comparable CPU evaluation, while SPEC CPU fails.

\subsubsection{The design of the experiment}
For CPU evaluatology, as discussed in Section~\ref{CPU_eva_four_steps}, we first constructed a billion-scale valid EC configuration space using all SPEC CPU2017 workloads and then sampled a subset of 32 EC configurations for each SPEC CPU2017 workload through iterative stratified random sampling. Using the subset of sampled EC configurations, we measured the execution time of one CPU under each EC configuration and calculated the execution time differences between two CPUs under equivalent EC configurations. After that, we calculated both the 95\% and 99\% confidence intervals of the execution time difference over the sampled configurations to establish the probabilistic bounds for population difference mean estimation. 

For the SPEC CPU2017 methodology, it recommended some component configuration while making other components undefined, details in Section~\ref{Sec-Mot}. We keep these undefined components the same with our constructed valid EC configurations in Section~\ref{CPU_eva_four_steps}. In practice, the industry users will use different configurations for those undefined components and hence will introduce many more biases.  We measured the execution time under this recommended configuration 3 times and chose the median for calculating the execution time difference between two CPUs.

Evaluating two CPUs using CPU evaluatology, we ensured that they were compared under equivalent EC configurations. However, for the SPEC CPU2017 methodology, the configurations applied to the two CPUs differed due to the different maximum number of hardware threads supported by CPUs (56 for x86 and 64 for ARM).

\subsubsection{Experiment result of one CPU}

\begin{figure}[h]
\centering
\includegraphics[scale=0.4]{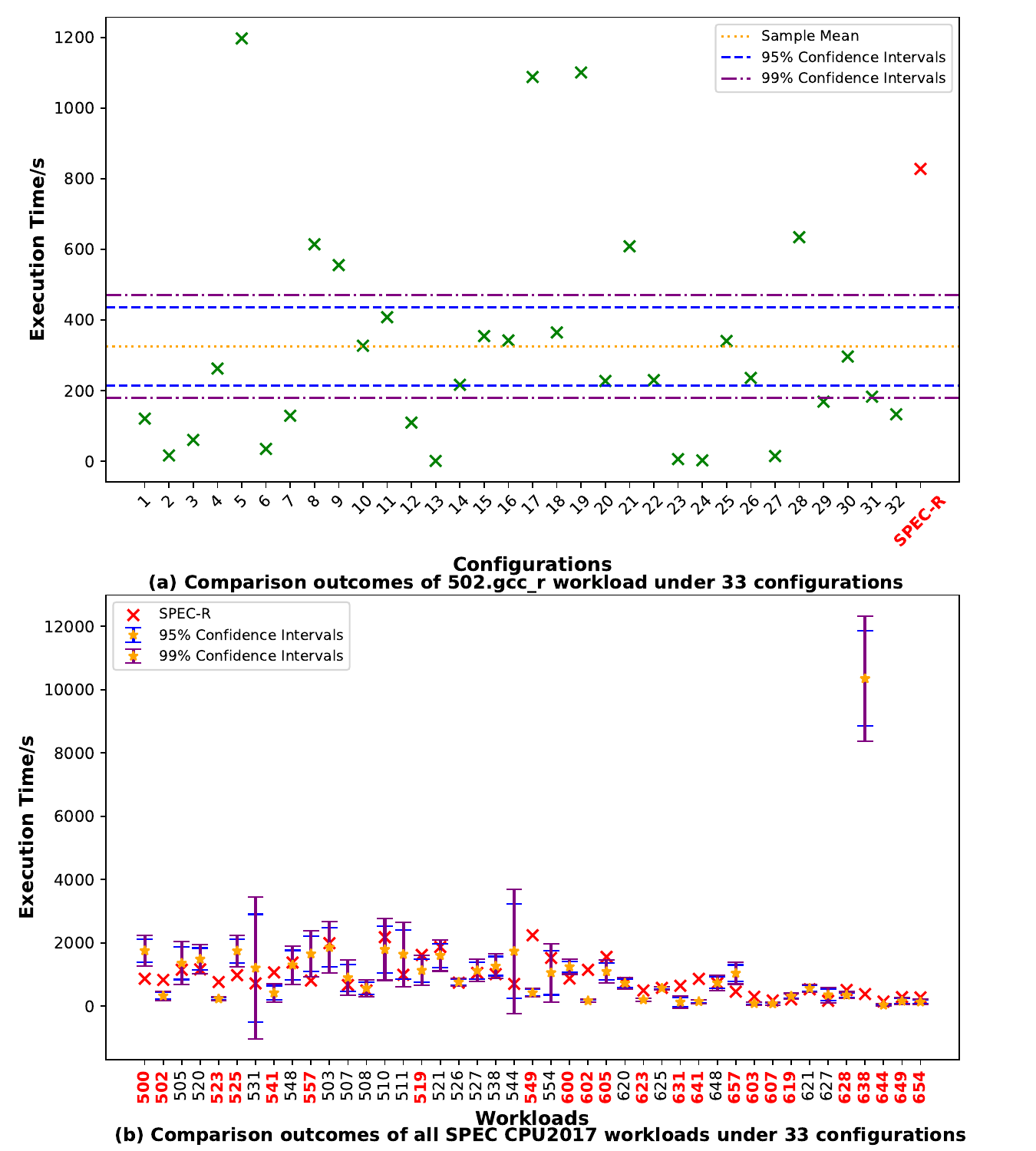}
\caption{Execution time of CPU A using CPU evaluatology against the SPEC CPU2017 methodology. (a) shows the Execution time of 502.gcc\_r workload using CPU evaluatology against the SPEC CPU2017 methodology. (b) shows the execution time of each SPEC CPU2017 workload using CPU evaluatology against the SPEC CPU2017 methodology.}
\label{SPEC-EC}
\end{figure}

Figure~\ref{SPEC-EC}(a) illustrates the execution time of CPU A under 33 different configurations using the 502.gcc\_r workload. The SPEC recommended configuration, labeled as `\textcolor{black}{\textbf{SPEC-R}},' is highlighted in red text, with its outcomes marked by red `\textcolor{black}{$\times$}' point on the plots. The outcomes from the remaining configurations are represented by green `\textcolor{green}{$\times$}' points. To gather this data, we conducted each configuration experiment three times and used the arithmetic mean as the outcome. In Figure~\ref{SPEC-EC}(a), a point (x, y) indicates that under the $x^{th}$ configuration, the execution time of the x86 processor running the 502.gcc\_r workload is y seconds. 
Additionally, we calculated the sample mean and both the 95\% and the 99\% confidence intervals. For the 502.gcc\_r workload, the sample mean of the evaluation outcome of CPU A is 324.625s, indicated by the orange dashed line, with a 95\% confidence intervals of 213.778s to 435.473s, marked by black dashed lines, \textcolor{black}{and a 99\% confidence intervals of 178.940s to 470.311s, marked by purple dashed lines}.

Figure~\ref{SPEC-EC}(b) presents I-beam plots summarizing similar data for all 43 SPEC CPU2017 workloads on CPU A. Each black I-beam represents the 95\% confidence intervals \textcolor{black}{and purple I-beam represents the 99\% confidence intervals} for the corresponding workload across 33 different configurations, while the orange star indicates the
sample mean. Please note that the I-beam plot labeled as 502 in Figure~\ref{SPEC-EC}(b) summarizes the data shown in Figure~\ref{SPEC-EC}(a) (the 502.gcc\_r workload). The red `\textcolor{black}{$\times$}' point indicates the execution time obtained using SPEC CPU2017 methodology. There are 23/43 workloads whose red `\textcolor{black}{$\times$}' points outside the purple I-beam. Thus, for these workloads, the execution time using SPEC CPU2017 methodology falls outside the 99\% confidence interval and is unrepresentative. Similarly, there are 27/43 workloads, whose execution time using SPEC CPU2017 methodology falls outside the 99\% confidence intervals on CPU B.

From Figure~\ref{SPEC-EC}, we draw two important points. First, 
\textcolor{black}{for a single workload, configuration variations can lead to substantial performance fluctuations, with execution times ranging from 1.129s to 1,197.155s (Figure~\ref{SPEC-EC}(a)). Across all workloads, the 99\% confidence intervals computed over 32 sampled configurations reveal highly variable widths — from as narrow as 67.588s to as wide as 4,466.190s (Figure~\ref{SPEC-EC}(b)) — demonstrating that different workloads exhibit differing sensitivities to configuration changes. Therefore, without maintaining consistent configurations for other components in the MES, we risk attributing performance differences—up to nearly 1000×—to the CPU itself, even when the CPU remains unchanged.}
\textcolor{black}{Second, the SPEC CPU2017 evaluation using the `-O3' compiler flag and core-matched threads holds no intrinsic significance — it is neither optimal, worst-case, average, nor median within the broader configuration space (Figure~\ref{SPEC-EC}(a)). As a single point in that space, the SPEC recommended configuration often falls outside the 95\% and 99\% confidence intervals (Figure~\ref{SPEC-EC}(b)), failing to reflect the wide distribution of user configurations and behaviors. This highlights the inadequacy of single-point evaluations and motivates our adoption of diverse configurations that reflect real-world user requirement distributions.}

\subsubsection{Experiment result of comparing two CPUs}

\begin{figure}[h]
\centering
\includegraphics[scale=0.4]{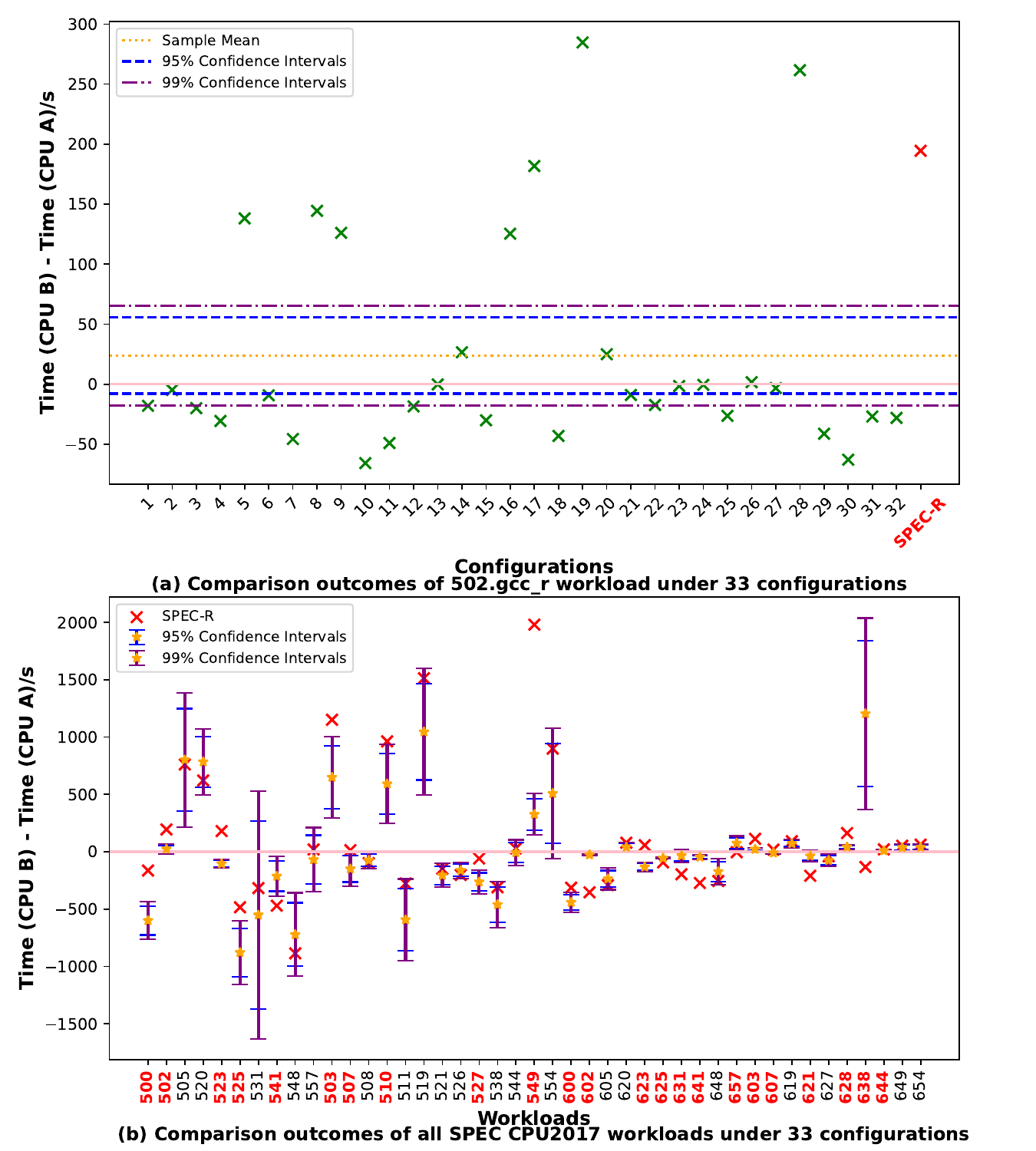}
\caption{The execution time differences between CPUs B and A using CPU evaluatology against the SPEC CPU2017 methodology. (a) shows the comparison outcomes of 502.gcc\_r workload, while (b) shows comparison outcomes of all SPEC CPU2017 workloads.}
\label{Method-SPEC-ARM-X86}
\end{figure}

We calculated the sample mean and both 95\% and 99\% confidence intervals of the execution time differences in Figure~\ref{Method-SPEC-ARM-X86} between the two CPUs under 32 sampled configurations through CPU evaluatology.  Figure~\ref{Method-SPEC-ARM-X86}(a) shows the sample mean, 95\% and 99\% confidence intervals of the execution time differences for the 502.gcc\_r workload under 32 sampled configurations. Across the 32 configurations, the sample mean of the execution time difference between the two CPUs is 23.888s, indicated by the orange dashed line, with 95\% confidence intervals ranging from -7.782s to 55.559s, marked by two black dashed lines, \textcolor{black}{and 99\% confidence intervals ranging from -17.736s to 65.512s, marked by two purple dashed lines}. This indicates that, on average, the execution time running the 502.gcc\_r workload on CPU A is 23.888s lower than that on CPU B. 
\textcolor{black}{The inclusion of 0 in the confidence interval indicates a lack of statistical significance, meaning no conclusive performance difference can be established between the two CPUs.}
However, the difference using SPEC CPU2017 methodology is 194.475s, indicating that CPU A outperforms CPU B in the 502.gcc\_r workload, which not only falls outside the 99\% confidence intervals but also contradicts the conclusion derived from the confidence intervals.

Figure~\ref{Method-SPEC-ARM-X86}(b) presents I-beam plots summarizing similar data for all 43 SPEC CPU2017 workloads. Each black I-beam represents the 95\% confidence intervals \textcolor{black}{and purple I-beam represents the 99\% confidence intervals} for the corresponding workload across the 32 sampled configurations, while the orange star indicates the sample mean. Please note that the I-beam plot labeled as 502 in Figure ~\ref{Method-SPEC-ARM-X86}(b) summarizes the data shown in Figure~\ref{Method-SPEC-ARM-X86}(a) (the 502.gcc\_r workload). The red `\textcolor{black}{$\times$}' point indicates the execution time difference obtained using SPEC CPU2017 methodology. There are 23/43 workloads (labeled in red text) whose red `\textcolor{black}{$\times$}' point falls outside the purple I-beam. Thus, for these workloads, the difference using SPEC CPU2017 methodology falls outside the 99\% confidence interval. 
Furthermore, for 10/43 workloads, the 99\% confidence intervals including 0 (blue dashed line) indicate no significant performance difference between CPU A and CPU B at the 99\% confidence level under the given sampled configurations.


\textcolor{black}{From Figure~\ref{Method-SPEC-ARM-X86}, we draw two important insights. First, confidence intervals offer statistically consistent conclusions across valid configurations, as all values within the interval are meaningful at the same confidence level. This ensures that the evaluation outcome is not overly influenced by any single configuration. Second, unlike the SPEC recommended configuration - which reflects only a single, arbitrarily ambiguous point in the configuration space, CPU evaluatology systematically explores a diverse set of configurations, providing a more comprehensive and controlled evaluation framework.
This is particularly critical when comparing different CPUs: relying solely on the SPEC methodology may yield biased or misleading conclusions, as its fixed configuration can either exaggerate or understate performance differences due to the other components unrelated to the CPU itself.}

\subsection{Comparing CPU evaluatology, SPEC CPU2017 methodology, two DOE methodologies and RCTs}\label{Mot_experiment_design}

As rigorous evaluation methodologies like the design of experiments (DOE) and Randomized Control Trials (RCTs) are widely used in other domains, some may argue that applying these methodologies to CPU evaluation could resolve empirical method limitations. 

In this subsection,  we systematically compared CPU evaluation methodologies, including CPU evaluatology, the SPEC CPU2017's practice, two canonical DOE methodologies, and RCTs. 

\subsubsection{The design of experiment}\label{Design_exp2}

Based on the 648.exchange2\_s workload, we follow the same methodology presented in Section~\ref{CPU_eva_four_steps} to construct a population of EC configurations. we use the population mean of evaluation outcomes as the ground truth to compare different methodologies.

For the design of the population of EC configurations,  our decision is as follows. 1) 10 validated datasets ($P_2$) (the 5 smallest datasets correspond to the low-level combinations in the $2^kr$ factorial design, the other 5 correspond to the high-level); 2) 3 compiler levels/flags ($P_4$) (`-O1'[low-level], `-O2', `-O3'[high-level]); 3) 24 expert-curated number of threads ($P_5$) (1, 2, 4, 8, 10, 16, 20, 30, 32, 40, 50, 56, 60, 64, 70, 80, 90, 100, 110, 120, 128, 200, 256, 300; the 12 smallest threads correspond to the low-level, the other 12 correspond to the high-level); 4) Keeping other system configurations (compiler[$P_3$], OS[$P_6$], OS setting[$P_7$], memory[$P_8$], memory setting[$P_9$], disk[$P_{10}$], and disk setting[$P_{11}$]) the same. We use the population mean of execution time differences between two CPUs across all 720 configurations as ground truth. 

For the $2^kr$ factorial and general factorial designs and RCTs, the design of EC configuration space relies upon the evaluator's expertise. To be fair, we reuse the expertise in CPU evaluatology.  The $2^kr$ factorial design chose k=3 factors: dataset, compiler level/flag, threads, 2 levels/factor, and r=3 replicates. The total configurations of evaluation are 24 ($8*3$). We calculated the 99\% confidence intervals for those 24 configurations.

For the general factorial design, we also chose k=3 factors: dataset, compiler levels/flags, threads, all factor levels, and 3 replicates. The total configurations of evaluation are 2160 ($720*3$). We calculated the 99\% confidence intervals.

For RCTs, we adopt without-replacement sampling with 3 replicates. we apply randomly sampled configurations to CPU A, and the other configurations of the same size to CPU B. We calculated the 99\% confidence intervals of the two independent sample mean differences. 

For CPU evaluatology, as discussed in Section~\ref{CPU_eva_four_steps}, we sampled a subset of 32 EC configurations through iterative stratified random sampling.

Given that experimental time costs linearly with the number of tested configurations, we employ the number of configurations evaluated per CPU as a proxy for the time costs metric. We defined the accuracy metric as coverage rates of 10,000 iterations where 99\% confidence intervals contained ground truth or the evaluation outcomes were consistent with the ground truth within an acceptable margin of error.

\begin{figure*}[h]
\centering
\includegraphics[scale=0.36]{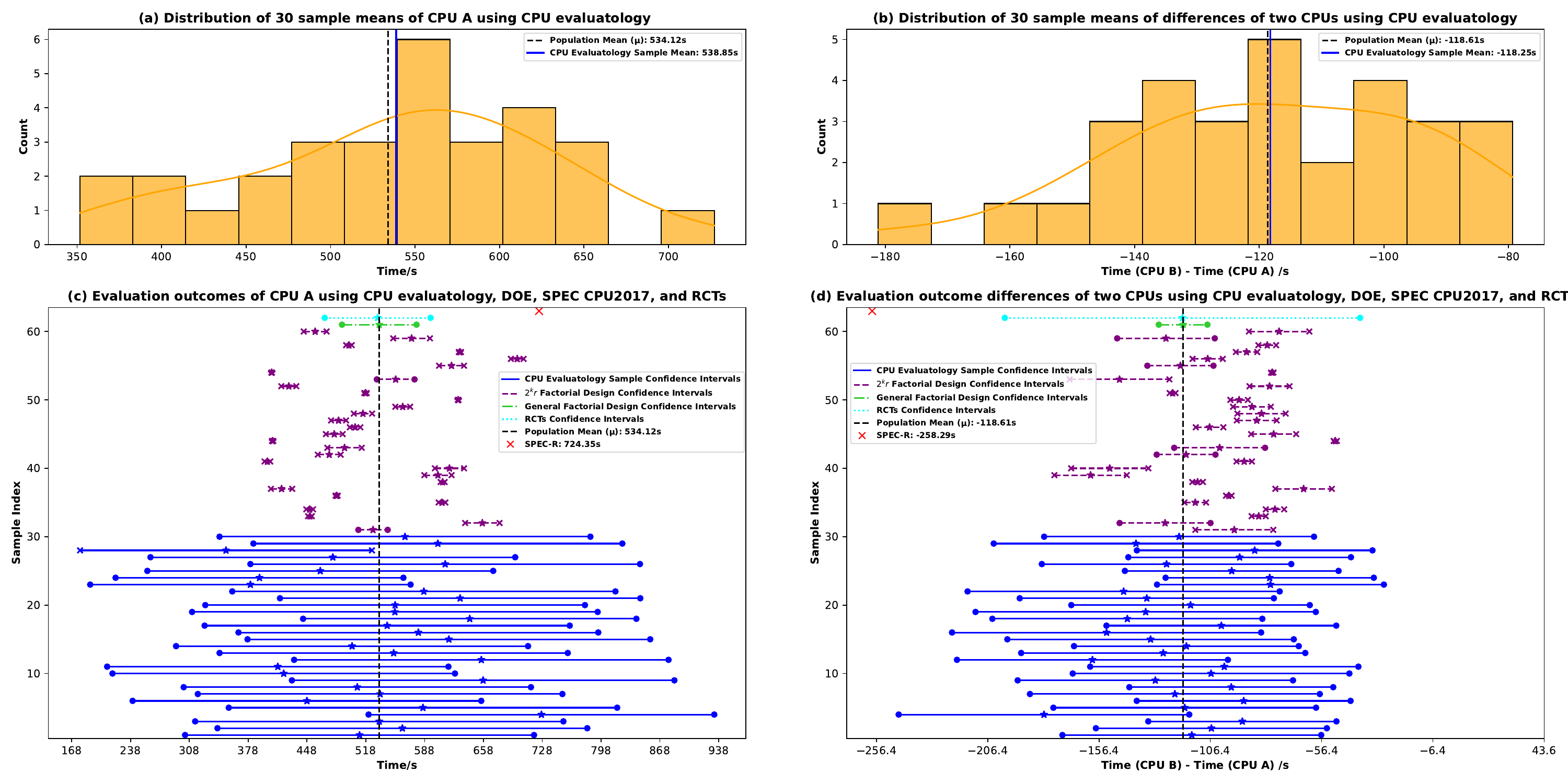}
\caption{
Comparisons among CPU evaluatology, $2^kr$ factorial and general factorial designs, SPEC CPU2017, and RCTs methodologies. All sub-figures use black dashed lines to denote the population mean (ground truth). (a) and (b) show the distribution of execution time sample means on CPU A and that of the difference between CPUs B and A using CPU evaluatology. (c) and (d) demonstrate evaluation outcomes being consistent toward the population mean for those methodologies on CPU A and between CPU B and CPU A. 
In (c) and (d): Samples 1-30 are \textcolor{black}{99\%} confidence intervals using CPU evaluatology. 31-60 are \textcolor{black}{99\%} confidence intervals from $2^kr$ factorial design. Sample 61 has \textcolor{black}{99\%} confidence intervals from the general factorial design. Sample 62 is the \textcolor{black}{99\%} confidence interval from RCTs. Sample 63 is the execution time using SPEC CPU2017 methodology.  
Confidence intervals with dot-shaped boundaries indicate inclusion of the ground truth, while those marked by crosses denote exclusion.
}
\label{Pop-com}
\end{figure*}

\subsubsection{Limitations of existing methodologies}

Table~\ref{Experiment_Design} presents the results. From this experiment, we can draw four important points. First, 
\textcolor{black}{owing to its reliance on a single recommended configuration and ambiguity in undefined component configurations, the SPEC CPU2017's methodology leads to substantial deviation from ground truth, achieving 0\% accuracy despite incurring minimal time cost.}
Second, the $2^kr$ factorial design exhibited unacceptably low accuracy, with only \textcolor{black}{25.10\%} of ground truth falling within the \textcolor{black}{99\%} confidence intervals, far below expectations. This stems from its severe undersampling: 8 tested configurations versus 720 possible combinations. Third, while the general factorial design achieved 100\% accuracy through exhaustive enumeration, its prohibitive time costs render it impractical for real-world applications, particularly in our billion-scale configuration space. Fourth, the RCTs' accuracy improved with sample size but demanded substantial costs; achieving \textcolor{black}{99\%$\pm$2\%} accuracy required 300 configurations per CPU (totaling 600), nearing the cost of full enumeration. Additionally, determining RCTs’ minimum sample size required iterative experiments, further inflating time costs. These findings highlight the dilemma of time costs and accuracy tradeoffs in existing CPU evaluation methodologies.

\begin{table}[]
\caption{The costs and accuracy (in 10,000 iterations) of different CPU evaluation methodologies.}
\begin{tabular}{lcl}
\hline
\textbf{Methodology}     & \textbf{Costs (Configs)} & \textbf{Accuracy} \\ \hline
SPEC CPU2017 & 1 per CPU                           & 0\%                  \\\hline
$2^kr$ Factorial Design  & 8 per CPU                           & 25.1\%                  \\
General Factorial Design & 720 per CPU                         & 100\%                    \\ \hline
RCTs                      & 150 per CPU                         & 90.00\%                  \\
RCTs                      & 200 per CPU                         & 94.10\%                  \\
RCTs                      & 250 per CPU                         & 97.50\%                   \\
RCTs                      & 300 per CPU                         & 97.30\%                  \\
RCTs                      & 350 per CPU                         & 99.10\%                   \\
RCTs                      & 360 per CPU                         & 99.40\%                  \\ \hline
CPU evaluatology                      & 32 per CPU                         & 97.32\%                  \\
CPU evaluatology                      & 50 per CPU                         & 97.91\%                  \\
CPU evaluatology                      & 60 per CPU                         & 98.38\%                  \\
CPU evaluatology                      & 80 per CPU                         & 98.28\%                  \\
CPU evaluatology                      & 90 per CPU                         & 98.36\%                  \\
CPU evaluatology                      & 150 per CPU                         & 98.51\%                  \\
CPU evaluatology                      & 200 per CPU                         & 98.84\%                  \\\hline
\end{tabular}
\label{Experiment_Design}
\end{table}

\subsubsection{Validation of consistent and comparable CPU evaluation outcomes with acceptable time costs in CPU evaluatology}\label{Validate_Com}

Figure~\ref{Pop-com} presents experimental results: left subplots show CPU A execution time results, while right subplots display CPU B minus CPU A execution time differences. CPU B results (omitted due to space constraints) exhibit patterns similar to CPU A. We conducted 30 sampling iterations for CPU evaluatology and $2^kr$ factorial design. The ground truths (black dashed lines) are 534.12s (CPU A population mean), 415.51s (CPU B population mean), and -118.61s (CPU B minus CPU A differences mean). Figure~\ref{Pop-com} (a) and (b) demonstrate CPU evaluatology's 30 sample mean distributions (blue solid lines represent the mean of CPU evaluatology sample means, purple dashed lines represent $2^kr$ factorial design mean, green dashed lines represent general factorial design mean, cyan dashed lines represent RCTs mean). The sample means in CPU evaluatology show consistent convergence toward the population mean, whereas DOE, RCTs, and SPEC CPU2017's practice exhibit larger deviation.
Figure~\ref{Pop-com} (c) and (d) compare those methodologies against ground truths. For CPU evaluatology, 29/30 (both CPU A and CPU B) and 30/30 (CPU B minus CPU A differences) \textcolor{black}{99\%} confidence intervals (dot-shaped boundaries) include the ground truth. For $2^kr$ factorial design, 2/30 (both CPU A and CPU B), and 5/30 (CPU B minus CPU A differences) 99\% confidence intervals include the ground truth. All the execution time using SPEC CPU2017 configurations (crosses-shaped boundaries) deviates from the ground truth. The 99\% confidence intervals of RCTs also include the ground truth.

From this experiment, There are three important points. First, due to the inherent limitations of 
the SPEC CPU2017 benchmark, all CPU evaluations using this benchmark fail to achieve consistent and comparable evaluation outcomes. Second, CPU evaluatology achieves consistent and comparable evaluation outcomes with acceptable time costs. While the general factorial design using CPU evaluatology yields the highest accuracy, its time costs are prohibitive. The RCTs, despite producing satisfactory accuracy, also incur excessively high time costs. Third, CPU evaluatology demonstrates robust stability in evaluation outcomes. Compared with methodologies and results in Table~\ref{Experiment_Design}, through 10,000 iterations, \textcolor{black}{98.25\% of 99\%} confidence intervals contained the population mean for CPU A, \textcolor{black}{98.29\%} for CPU B, and \textcolor{black}{97.32\%} for CPU B minus CPU A execution time differences. Furthermore, with sample sizes $\geq$ 30 (large-sample criteria), the coverage rates of confidence intervals fluctuate near the \textcolor{black}{99\%} confidence level. And increasing the sample size beyond 30 yields no significant improvement, with \textcolor{black}{98\%} coverage rates at a sample size of 60, stabilizing around \textcolor{black}{99\%} from a sample size of \textcolor{black}{200}. Considering the tradeoff of time costs and accuracy, we use the sample size of 32 for CPU evaluatology.

\section{Conclusion}\label{Sec_Conclusion}

\textcolor{black}{This article reveals unique challenges of CPU evaluation: user-perceived performance metrics can only be measured on a minimal evaluation system (MES) consisting of the CPU and other indispensable components (evaluation condition). The challenge of the CPU evaluation is to accurately attribute the deviations in the evaluation outcomes on the MES to the differences between the CPUs and avoid the confounding from the configuration of the evaluation condition.}

\textcolor{black}{Our experiments reveal that the industry-standard CPU benchmark, SPEC CPU2017, suffers from a significant flaw: for the identical CPU, undefined configurations of evaluation conditions introduce uncontrolled variability in evaluation outcomes, and the confounding from the evaluation condition significantly distorts the evaluation outcomes.}

\textcolor{black}{To bridge this gap, we formally define the CPU evaluation problem and introduce a rigorous CPU evaluation methodology--CPU evaluatology. We demonstrate that CPU evaluatology yields consistent and comparable evaluation outcomes with acceptable time costs, outperforming SPEC CPU2017, two DOE, and one RCTs methodologies.}



\bibliographystyle{ACM-Reference-Format}
\bibliography{sample-base}

\end{document}